\documentclass[conference]{IEEEtran}

\usepackage{amsmath,amssymb,amsfonts}
\usepackage{amssymb,mathtools}
\usepackage{algorithm,algpseudocode}
\usepackage{enumitem}
\usepackage{graphicx}
\usepackage{textcomp}
\usepackage{xcolor}
\usepackage{xspace}
\usepackage{todonotes}
\usepackage{enumitem}
\usepackage{url}
\usepackage{hyperref}
\usepackage{subcaption}
\usepackage[export]{adjustbox}
\usepackage{flushend}
\usepackage[T1]{fontenc}

\newcommand{\aagent}{\texttt{active agent}\xspace}
\newcommand{\pagent}{\texttt{passive agent}\xspace}
\newcommand{\approach}{\texttt{cMarlTest}\xspace}
\newcommand{\tool}{\texttt{RLbT}\xspace}

\newcommand{\rl}{reinforcement learning\xspace}
\newcommand{\RL}{Reinforcement Learning\xspace}
\newcommand{\MARL}{Multi-Agent Reinforcement Learning\xspace}
\newcommand{\marlacronym}{MARL\xspace}
\newcommand{\rlacronym}{RL\xspace}

\newcommand{\sut}{\texttt{Lab Recruits}\xspace}
\newcommand{\lr}{\texttt{Lab Recruits}\xspace}

\newcommand{\buttonsdoors}{\texttt{L1}\xspace}
\newcommand{\med}{\texttt{Small}\xspace}
\newcommand{\lrg}{\texttt{Medium}\xspace}
\newcommand{\ext}{\texttt{Large}\xspace}

\newcommand{\rqa}{\textbf{RQ1}\xspace}

\newcommand{\rqb}{\textbf{RQ2}\xspace}

\begin{document}

\title{Curiosity Driven Multi-agent Reinforcement Learning for 3D Game Testing}

\author{\IEEEauthorblockN{Raihana Ferdous}
\IEEEauthorblockA{\textit{Consiglio Nazionale delle Ricerche} \\
Pisa, Italy \\
raihana.ferdous@isti.cnr.it}
\and
\IEEEauthorblockN{Fitsum Kifetew}
\IEEEauthorblockA{\textit{Fondazione Bruno Kessler} \\
Trento, Italy \\
kifetew@fbk.eu}
\and
\IEEEauthorblockN{Davide Prandi}
\IEEEauthorblockA{\textit{Fondazione Bruno Kessler} \\
Trento, Italy \\
prandi@fbk.eu}
\and
\IEEEauthorblockN{Angelo Susi}
\IEEEauthorblockA{\textit{Fondazione Bruno Kessler} \\
Trento, Italy \\
susi@fbk.eu}
}

\maketitle

\begin{abstract} 
Recently testing of games via autonomous agents has shown great promise in tackling challenges faced by the game industry, which mainly relied on either manual testing or record/replay. In particular \RL (\rlacronym) solutions have shown potential by learning directly from playing the game without the need for human intervention.

In this paper, we present \approach, an approach for testing 3D games through curiosity driven \MARL (\marlacronym). \approach deploys multiple agents that work collaboratively to achieve the testing objective.
The use of multiple agents helps resolve issues faced by a single agent approach.

We carried out experiments on different levels of a 3D game comparing the performance of \approach with a single agent \rlacronym variant. Results are promising where, considering three different types of coverage criteria, \approach achieved higher coverage. \approach was also more efficient in terms of the time taken, with respect to the single agent based variant.

\end{abstract}


\section{Introduction} \label{sec:introduction}
Software testing has long been recognized as a critical aspect of the development process and accounts for a significant portion of the overall budget. Over the years researchers have developed numerous techniques and tools to automate as much as possible the testing process, hence reducing the associated cost. Despite research efforts, testing remains a primarily manual task for certain software systems, such as computer games~\cite{DBLP:conf/ast/PolitowskiPG21}. Computer games, and particularly 3D games, bring several challenges to the table when considering their automated testing. They are highly interactive, involving continuous input spaces, as well as visual elements that are not easy to deal with in an automated manner. Consequently, automated testing of games has been mostly limited to capture-replay based solutions, which tend to produce brittle tests. Recent research using Reinforcement Learning (\rlacronym) has shown promising results~\cite{silver2017masteringGo,mnih2013playing,openAI-dota2,DBLP:journals/tciaig/AlbaghajatiA23,politowski2022towards, DBLP:journals/tciaig/AspertiCPPS20},
where \rlacronym is used to train agents that can play the game and win it.  There have been recent efforts in the automated testing of the wider scope of software applications that fall under the category of Extended Reality (XR) systems~\cite{DBLP:conf/icst/PradaPKDVLDKDF20}, which include 3D games as well.

 In a previous work, we proposed using single agent \rlacronym for testing games where we obtained encouraging results~\cite{DBLP:conf/kbse/FerdousKPS22}. We also observed that while \rlacronym works reasonably well and serves as a good starting point, it faces a number of challenges that are difficult to tackle with a single agent. One such challenge is the fact that modern 3D games are only partially observable. Hence, a single agent faces difficulty in properly assessing the effect(s) of its actions. For instance, at a certain point in the game, if the agent flips a light switch that controls several light bulbs distributed in different rooms/corridors (some of which could be behind closed doors), it is extremely difficult for the agent to observe the effect of the switch (i.e., which bulbs are alight and which not). Similarly, the size of the game is also another challenge where the agent would need to cover a lot of ground to fully explore the game level~\cite{DBLP:conf/kbse/FerdousKPS22}.

To address the aforementioned challenges, in this paper, we propose \approach, a testing approach for 3D games  exploiting \marlacronym where two (or more) agents work collaboratively to explore the game world and achieve the desired testing objectives. \approach uses a curiosity driven reward mechanism in such a way that the agents are driven towards unexplored areas of the game world, maximizing coverage of the game. This work builds on our previous work~\cite{DBLP:conf/kbse/FerdousKPS22} using a similar setup and environment but instead, it applies a multi-agent \rlacronym approach to address some of the limitations of the aforementioned work. The use of a similar setup and environment enables a direct comparison of the results of the single agent and multi agent \rlacronym in the same context.

This paper presents details about \approach and the experiment we carried out to assess its performance with respect to a single agent \rlacronym based approach. 
Specifically, we aim to answer the following research questions:

\noindent 
\textbf{\rqa (Effectiveness)} What is  the effectiveness of \approach for testing 3D games with respect to the single agent \rlacronym approach? Effectiveness is measured by calculating the coverage of the 3D game achieved by the compared approaches.

\noindent
\textbf{\rqb (Efficiency)} What is  the efficiency of \approach for testing 3D games with respect to the single agent \rlacronym approach? Efficiency is measured by calculating the amount of time required by the compared approaches to achieve their final coverage. 

The rest of the paper is organized as follows: in Section~\ref{sec:runningexample} we introduce a running example we use in the rest of the paper. In Section~\ref{sec:approach} we present \approach, followed by the experimental setup in Section~\ref{sec:experiment}. The results of the experiment are presented in  Section~\ref{sec:results}. Section~\ref{sec:relatedworks} discusses related works, and finally Section~\ref{sec:conclusion} concludes the paper and outlines future work.

\section{Running Example} \label{sec:runningexample}

\begin{figure}[tb]
    \begin{center}
    \includegraphics[width=0.48\textwidth]{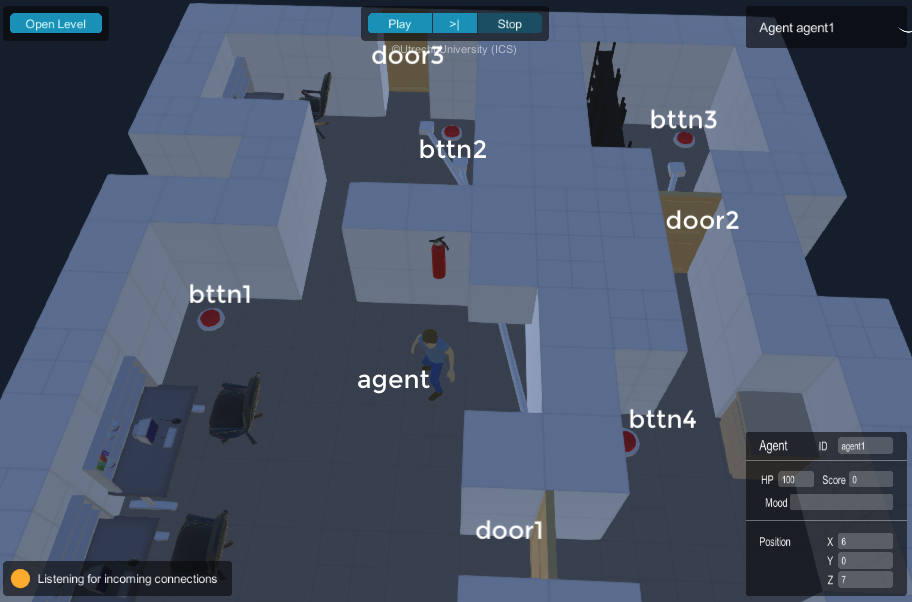}
    \end{center}
    \caption{Level \buttonsdoors~in \sut.} 
    \label{fig:runningexample}
\end{figure}

This section introduces an example taken from \sut, an open source 3D game\footnote{\url{https://github.com/iv4xr-project/labrecruits}} where the player navigates a maze-like environment composed of rooms and corridors connected by doors. The doors are connected to buttons that open/close one or more doors. 
The game level may also contain other entities, such as 
desks, chairs, fire hazards, extinguishers, etc. \sut allows the user to design and load a game level. Our example, illustrated in Figure~\ref{fig:runningexample}, is a simple level in \sut composed of three doors (door1, door2, door3), four buttons (bttn1, bttn2, bttn3, bttn4), and other non interactable entities such as desks. \sut could be played by either human players or test agents~\cite{prasetya2020aplib} by controlling the avatar (labelled as `agent' in Figure~\ref{fig:runningexample}). Our approach uses the test agents for interacting with the game.

\section{\approach: Curiosity Driven \marlacronym for Game Testing} \label{sec:approach}
Our proposed approach is to employ \marlacronym where multiple agents work collaboratively to achieve better exploration of the game world. In the literature, \marlacronym based approaches have been applied in the context of game play~\cite{OpenAIHideandSeekbaker2019,vinyals2019grandmasterStarCraft,park2019multi,zhou2021hierarchical}, where the main objective is typically to train agents that can play and eventually win the game. 
In our case the objective is not to train agents capable of winning a game, but rather agents that can effectively explore the game world from the perspective of \emph{software testing} where the ultimate goal is to achieve full coverage of the desired coverage criterion. Different notions of coverage could be considered, depending on the game and desired testing level~\cite{10.1145/3643658.3643920}. In our experiments, we consider three coverage criteria, i.e., \emph{entity coverage}, \emph{entity connection coverage}, and \emph{spatial coverage} (as detailed in Section~\ref{sec:metrics}). However, the approach can easily be extended to include other coverage criteria~\cite{10.1145/3643658.3643920}.

Deployment of the \marlacronym architecture for testing of 3D games is  difficult~\cite{canese2021multi}, 
one of the prominent challenges being the non-stationarity of the environment. 
Though \marlacronym agents are autonomous entities with individual goals and independent decision making capabilities, they are influenced by each other's decisions while simultaneously learning by interacting with a shared environment. 
This indicates that the environment dynamics become non-stationary making learning an optimal policy difficult. 
Non-stationarity of the environment violates the Markov property which is the foundation of the basic \rlacronym approach stating that the environment where the agent is learning is stationary, and, the state transition and reward depend only on the current state of an agent~\cite{sutton2018reinforcement}. Hence, a Markov Decision Process \cite{bellman1957markovian} is no longer feasible to describe \marlacronym. Thus, \marlacronym which considers the decision making process involving multiple agents is modelled through a stochastic game, also known as a Markov game \cite{littman1994markov}. 

Various approaches and frameworks are proposed in recent research to solve the Markov game \cite{DBLP:journals/corr/abs-2011-00583, zhang2021multi} and to handle non-stationarity issues \cite{DBLP:journals/corr/abs-1906-04737}. These approaches range from using modifications of standard \rlacronym training methods to computing and sharing additional information. 
Although these policies can moderate the non-stationarity of the environment, learning algorithms still suffer from instability, 
especially in games where agents only partially observe the environment.
For example, the problem of testing and providing coverage of 3D games which we formulate as a cooperative multi-agent \rl problem becomes a particularly challenging class of problem due to the presence of partial observability. Here, agents must learn to coordinate in a non-stationary environment, while relying only on their partial information/observations. The agents may be oblivious to the actions of other agents even though these actions have a direct impact on the environment, and hence on their reward. 

One possible solution to address the partial observability issue in the cooperative \marlacronym is the possibility of sharing experience between the agents \cite{gerstgrasser2024selectively}. 
Here, we have proposed \approach, a cooperative \marlacronym solution where we define the agents to be fully or partially independent with different roles, meaning they are assigned to achieve different but related sub-goals. Performing different sub-tasks ensures that they are independent, hence, they can be benefited by sharing experience to achieve the main goal that is, to maximize game coverage. 
The rest of this section describes the details of our proposed approach \approach for \marlacronym-based testing of 3D games.   

\subsection{States and actions}
\approach relies on the underlying agent-based testing platform~\cite{prasetya2020aplib} for interacting with the game under test (GUT). The underlying test agent can perform low-level activities in the game world with complete autonomy. For instance, the test agent can navigate from point A to point B in the game on its own, where points A and B are identified by entities in the game (e.g., 
 doors) rather than pixel level coordinates as is typical in game testing contexts. This enables \approach to remain at a higher level of abstraction in terms of the states and actions available to the \rlacronym agent. In other words, when formulating the \rlacronym problem, we can define high-level states and actions rather than low-level ones. An example of a high-level action could be \emph{go to door A}, while a state could be \emph{door A is open, door B is closed, and button C is pressed}. Such state/action specifications are high-level with respect to low level actions such as \emph{move left for 10 pixels}. \approach uses the agent library of the iv4XR framework which provides a high-level view of the agent's observations of the environment~\cite{prasetya2020aplib}.

The states in \approach are defined in terms of game entities and their attributes. Let $E$ be the set of entity types in a game $G$ and 
$Attr_{E_i}$ the set of attributes of entity type $E_i$,
where each attribute $Attr$ can assume a set of values $Val$. The set of possible states is defined as: $$E \times Attr \times Val$$ 
In the case of \sut, we have
\begin{itemize}
    \item $E = \{Door, Button\}$
    \item $Attr_{Door} = \{ID, isOpen\}$ \\ $Attr_{Button} = \{ID, isPressed\}$
    \item $Val_{isOpen} = Val_{isPressed} = \{true, false\}$
\end{itemize}
Hence, for a given level of \sut, the state space is defined by the combination of the values of the doors and buttons in the level. For instance, the level \buttonsdoors in our running example (see Figure~\ref{fig:runningexample} in Section~\ref{sec:runningexample}) includes three doors and four buttons and  the possible states are \{(door1, isOpen, true), (door1, isOpen, false), (door2, isOpen, true), (door2, isOpen, false), (door3, isOpen, true), (door3, isOpen, false), (bttn1, isPressed, true),(bttn1, isPressed, false), (bttn2, isPressed, true), (bttn2, isPressed, false), (bttn3, isPressed, true), (bttn3, isPressed, false), (bttn4, isPressed, true), (bttn4, isPressed, false)\}.

At a given state $S$ of the game, the possible set of actions are those allowed by the game for the entities available in $S$. The specific set of actions depends on the GUT. For instance, in the case of \sut, typical possible actions include pressing a button, going through a door, etc. Going back to our running example, given a state $S = (bttn2, isPressed, true), (door3, isOpen, false)$ (this corresponds to a situation where the player is near the button \emph{bttn2}), the possible actions are either to interact with button \emph{bttn2} or to interact with (go through) door \emph{door3}. Note that depending on the position of the player in the game, the observed state is typically partial, and hence the set of possible actions is also limited to those entities available in the current state.

\subsection{Agent roles}
\approach uses a \marlacronym setup in which multiple agents work collaboratively to achieve full \emph{coverage} of the desired adequacy criterion. 
In general, \marlacronym agents could be cooperative, competitive, or a mix~\cite{DBLP:journals/tsmc/BusoniuBS08}. While in this work we adopt cooperative agents, other types of multi agent configurations, e.g., competitive agents, could be further explored as part of future studies. 

In the context of this paper, the testing objective is to explore the game world until some adequacy criterion (coverage) is satisfied. To this end, we adopted a fully cooperative \marlacronym scheme in which agents with distinct roles work together to achieve \emph{full coverage}.  Specifically, \approach defines two agent roles: \aagent and \pagent. The \aagent is responsible for actively interacting with the environment by performing \emph{actions} to learn an optimal policy, while the \pagent is limited to only observing the environment and sharing the observation with the \aagent, without performing any actions. The \aagent behaves as a typical \rlacronym agent that observes the environment and performs actions, with the only addition that its observation is further \emph{enriched} with the observations it receives from one or more {\pagent}s. On the other hand, the \pagent performs only observations of the environment and sends them to the \aagent. 

\subsection{Reward function} \label{subsec:rewardfunction}
\approach employs a reward function that promotes the exploration of new areas in the game, hence the agents are encouraged to follow their curiosity \cite{pathak2017curiosity}, and discouraged from revisiting previously seen areas and from being stationary.
We formulate the reward proportional to the transition's novelty, providing a low and decreasing reward (e.g., penalty) for revisiting previously explored states, while a high reward is given for reaching new areas or triggering new actions.
The reward is computed based on how distant the current observation is from the most similar observation in memory using the Jaccard coefficient \cite{jaccard1912distribution}. The reward mechanism is similar to the one employed in our previous work with single agent \rlacronym~\cite{DBLP:conf/kbse/FerdousKPS22}.

\subsection{Methodology}
The two types of agents defined in \approach are activated simultaneously. 
After executing an action, the \aagent observes the current state of the environment according to its observation range. The \pagent aids the \aagent to improve this knowledge by sharing the information it gathers about the current state of the environment from its current location in the game world. Note that the observations of the two agents could be identical (if they are close to one another), completely different (if they are too far apart in different areas of the game world), or partially overlapping depending on their locations.

The \aagent uses Q-learning, a model-free \rlacronym algorithm that does not require prior knowledge of the environment and is applicable in different environments. It is a value-based, off-policy algorithm that tries to find the best series of actions based on the agent's current state. 
The curiosity-based reward mechanism (described in \ref{subsec:rewardfunction}) is used to calculate the reward corresponding to an action chosen by the agent. A tabular Q-learning solution is used to keep the design simple. To ensure convergence and reduce the Q-table size, a state similarity measure is adopted when updating the Q-table. Instead of making a new entry for every state-action pair, a similarity calculation is performed to identify the most similar state, if it exists, in the Q-table. 

\emph{Decayed Epsilon-Greedy} is used as the initial policy to balance between  exploration and exploitation by allowing the \aagent to explore more when it does not have enough information about the environment, and to do exploitation once it has gathered enough information by interacting with the environment. 
The exploration-exploitation  process is controlled by the $\epsilon$ parameter, which \emph{is decayed} by a factor in each episode. The decaying factor is calculated based on the number of learning episodes.

\section{Evaluation Setup} \label{sec:experiment}
We have implemented \approach in a prototype tool and carried out experiments in order to assess its suitability for testing 3D games with respect to a single agent \rlacronym implementation. In this section, we present the setup of the experiment, the metrics used, the various artefacts, and the objects of the experiment. Subsequently, in Section~\ref{sec:results}, we discuss the results obtained and answer the research questions.

\subsection{Prototype} \label{sec:prototype}
\approach is implemented in a prototype tool called \tool. The implementation of \tool is built on top of the iv4XR framework for automated testing of extended reality (XR) based systems~\cite{DBLP:conf/icst/PradaPKDVLDKDF20}. \tool provides an implementation of \approach as described in Section~\ref{sec:approach} as well as an implementation of the single agent variant~\cite{DBLP:conf/kbse/FerdousKPS22}, which we use as a baseline. \tool is developed in Java and uses the BURLAP\footnote{\url{http://burlap.cs.brown.edu/}} library for \rlacronym algorithms. It is fully open source and publicly available on GitHub\footnote{
\url{https://github.com/iv4xr-project/iv4xr-rlbt}}.

\subsection{Game levels}
\label{sec:levels}
For the experiment, we selected three configurations of \sut, each representing a different scenario with increasing size and complexity:
\med, \lrg, and \ext. For each configuration, we prepared five different levels to have different levels of comparable difficulty but different layouts of the game world and entities. The difficulty faced by the testing approach results from the physical dimension of the level, the number of entities in it, and the connections among the entities which make navigation increasingly difficult.

\subsection{Metrics} \label{sec:metrics}
To answer our research questions, we measure coverage achieved by the testing approaches and the corresponding time spent. With respect to \emph{effectiveness} (\rqa), we consider three different notions of coverage related to gameplay~\cite{10.1145/3643658.3643920} for measuring the suitability of the proposed approach in exploring the game under test: \emph{entity coverage}, \emph{entity connection coverage}, and \emph{spatial coverage}.

\textbf{Entity coverage} represents the percentage of observed/interacted entities (with all possible property values) in a given level of \sut. In our running example, the \buttonsdoors level (shown in Figure~\ref{fig:runningexample}) contains seven entities, i.e., three doors and four buttons. A door can be in two states, i.e., \emph{open} or \emph{closed}, and similarly, a button could be in two states, i.e., \emph{pressed} or \emph{not-pressed}. Hence the total space of entities to be covered in a level consists of all these possible combinations. Table~\ref{tab:entitycoverage} lists all the possible entity states for the \buttonsdoors level shown in Figure~\ref{fig:runningexample}.

{
\begin{table}[!htb]
    \centering
    \begin{tabular}{c|c}
        \textbf{Entity} & \textbf{Properties}  \\ \hline
        \texttt{bttn1} & \it{pressed}, \it{not-pressed}\\
        \texttt{bttn2} & \it{pressed},  \it{not-pressed}\\
        \texttt{bttn3} & \it{pressed},  \it{not-pressed}\\
        \texttt{bttn4} & \it{pressed},  \it{not-pressed}\\
        \texttt{door1} & \it{open},  \it{closed}\\
        \texttt{door2} & \it{open},  \it{closed}\\
        \texttt{door3} & \it{open},  \it{closed}\\
        \hline
        \end{tabular}
    \caption{Coverable entities for the running example (see Figure~\ref{fig:runningexample})}
    \label{tab:entitycoverage}
\end{table}
}

\textbf{Entity connection coverage} represents the percentage of the connections exercised among the various entities in a level. In \sut, doors are connected to buttons in such a way that when a button is toggled it opens/closes the door(s) it is connected to. The entity connection coverage metric measures what percentage of such connections is exercised during testing. For our running example, there are five connections to be covered: \texttt{bttn2} is connected to \texttt{door1}, \texttt{bttn3} is connected to \texttt{door1, door2, door3}, and \texttt{bttn4} is connected to \texttt{door1}. Button \texttt{bttn1} is not connected to any doors, as shown in Table~\ref{tab:connectioncoverage}.

 \begin{table}[!h]
     \centering
     \begin{tabular}{c|c}
         \textbf{Buttons} & \textbf{Connecting Doors}  \\ \hline
         \texttt{bttn1} & \it{not connected}\\
         \texttt{bttn2} & \it{door1}\\
         \texttt{bttn3} & \it{door1}, \it{door2}, \it{door3}\\
         \texttt{bttn4} & \it{door1}\\
         \hline
         \end{tabular}
     \caption{Entity connections (see Figure~\ref{fig:runningexample})}
     \label{tab:connectioncoverage}
 \end{table}

\textbf{Spatial coverage} represents a visual representation of the exploration achieved by the agent (player) on a 2D plane using a heat map. 

With respect to \emph{efficiency} (\rqb), we measure the time spent by the approach to perform the testing. A testing run terminates either when all coverable goals have been covered (i.e., 100\% coverage is reached), when the allocated budget runs out (i.e., the number of episodes), or when the game play terminates (e.g., the player dies\footnote{In the case of \sut a player dies when the health point reaches~0 due to repeated travel through fires.}). Specifically, we calculate the \emph{average time per episode} as well as the \emph{average time per action} in a given episode (episode time divided by the number of actions performed in the episode).

\subsection{Experiment protocol} \label{sec:experimentexecution}
We run both \approach and the baseline (i.e., single-agent variant) in a similar environment and implementation to minimize any differences between the two approaches other than the \rl strategy. We run both variants on each level of \lr selected for the experiment (Section~\ref{sec:levels}). Each run was repeated 10 times to account for the inherent randomness in the approaches, and eventually, the metrics were computed as average over the 10 runs. We use the Wilcoxon test in order to determine statistical significance when comparing the performances of the two variants. 

\tool has several parameters that control different aspects of the testing process. Some of the parameters are specific to the game under test while others are specific to \rl aspects. After some preliminary experimentation, we have chosen reasonable values for the parameters to be used in our experiment. Table~\ref{tab:parameters} lists some of the important parameters and their corresponding values. Since the game levels used in our experiments are of different sizes, some of the parameter values are different among the levels, for instance, the number of \emph{cycles per action}, as can be seen in Table~\ref{tab:parameters}. This parameter represents the maximum number of cycles that the agent could execute to complete a given action. The parameter refers to the underlying execution environment which is based on BDI (belief-desire-intention) agents that execute in update cycles~\cite{DBLP:conf/icst/PradaPKDVLDKDF20}. To execute an action selected by the \rlacronym agent in the game (say to move from the current position to a given door), the underlying environment needs to execute for several update cycles in which the test agent repeatedly observes the world and moves towards the destination. If the game world is large, the agent might need more time to perform an action as compared to a smaller game world. Accordingly, we used different values for each level size in our experiment, i.e., 70, 100, and 120 cycles per action for \med, \lrg, and \ext respectively.

\begin{table}[!htb]
\centering
\begin{tabular}{l|l|l|l}
\hline \textbf{Parameter}  & \begin{tabular}[c]{@{}l@{}}\textbf{\med}\\\textbf{Levels}\end{tabular} & \begin{tabular}[c]{@{}l@{}}\textbf{\lrg}\\\textbf{Levels}\end{tabular}  & \begin{tabular}[c]{@{}l@{}}\textbf{\ext}\\\textbf{Levels}\end{tabular}\\ \hline
    num. agent & 2 & 2  & 2  \\
    num. episodes & 50 & 50  & 50  \\
    actions per episode & 80 & 150 & 200\\
    cycles per action & 70 & 100 & 120 \\
    initial $\epsilon$-value & 0.5 & 0.5 & 0.5 \\
    learning rate ($\alpha$) & 0.25 & 0.25 & 0.25\\
    discount rate ($\gamma$) & 0.6 & 0.6 & 0.6\\
    \hline
    \end{tabular}
    \caption{Parameter values}
    \label{tab:parameters}
\end{table}

\section{Results and Discussion} \label{sec:results}

\begin{figure*}[!htb]
\includegraphics[width=1\textwidth]{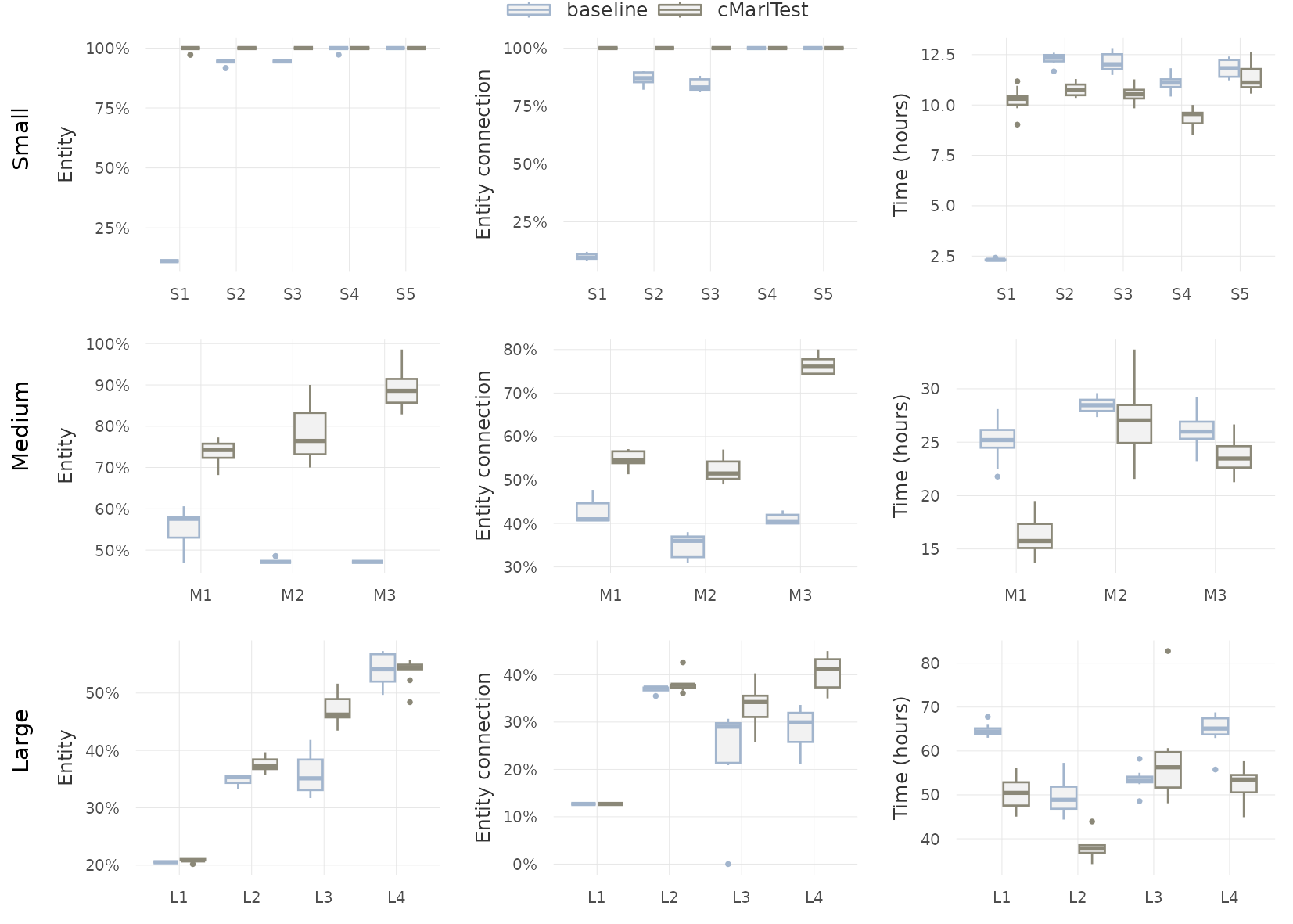}
\caption{Entity coverage, entity connection coverage and time boxplots}
 \label{fig:Coverage-and-Time}
\end{figure*}

\begin{table*}[!htb]
{
\begin{center}
\begin{tabular}{|ll|lll|lll|lll|}
\hline
 &    & \multicolumn{3}{|c|}{Entity coverage} & \multicolumn{3}{|c|}{Entity  connection coverage} & \multicolumn{3}{|c|}{Time} \tabularnewline
 Size & Id & P-value & Effect size & Magnitude & P-value & Effect size & Magnitude & P-value & Effect size & Magnitude \tabularnewline
\hline
\hline
Small&S1&\textless 0.0001&$1.000$&large&\textless 0.0001&$1.000$&large&0.00018&$1.000$&large\tabularnewline
Small&S2&\textless 0.0001&$1.000$&large&\textless 0.0001&$1.000$&large&0.00028&$0.000$&large\tabularnewline
Small&S3&\textless 0.0001&$1.000$&large&\textless 0.0001&$1.000$&large&0.00018&$0.000$&large\tabularnewline
Small&S4&0.37&NA&NA&NA&NA&NA&0.00018&$0.000$&large\tabularnewline
Small&S5&NA&NA&NA&NA&NA&NA&0.076&NA&NA\tabularnewline \hline

Medium&M1&\textless 0.0001&$1.000$&large&0.00017&$1.000$&large&\textless 0.0001&$0.000$&large\tabularnewline
Medium&M2&0.00011&$1.000$&large&0.00018&$1.000$&large&0.054&NA&NA\tabularnewline
Medium&M3&\textless 0.0001&$1.000$&large&0.00015&$1.000$&large&\textless 0.0001&$0.135$&large\tabularnewline \hline

Large&L1&0.0012&$0.900$&large&NA&NA&NA&0.00018&$0.000$&large\tabularnewline
Large&L2&0.00026&$0.985$&large&0.033&$0.775$&large&0.00018&$0.000$&large\tabularnewline
Large&L3&0.0014&$1.000$&large&0.0017&$0.920$&large&0.36&NA&NA\tabularnewline
Large&L4&0.91&NA&NA&0.00018&$1.000$&large&0.00025&$0.010$&large\tabularnewline
\hline
\end{tabular}\end{center}
}
\caption{Summary statistics}
\label{tab:Statistics}
\end{table*}

This section presents the results achieved by \approach and the baseline on each of the subjects of the experiment according to the metrics defined in the previous section. For effectiveness (\rqa) we report the coverage values achieved by the approaches for \emph{entity} and \emph{entity connection} coverage metrics. For \emph{spatial} coverage instead, we present only two heatmaps, because of space limitations. However, we make available an online appendix~\cite{online-appendix} with all data. For efficiency (\rqb) we report the time spent for each approach on the different levels of the game under test.

Figure~\ref{fig:Coverage-and-Time} shows the entity coverage for all the \lr levels in the experiment and the total time spent. Please note that one level for \med and two levels for \ext did not finish the executions, hence the plots reported in the figure are missing these three data points. Table~\ref{tab:Statistics} presents the results of the Wilcoxon test of statistical significance, both for coverage and time, as well as the effect size computed using the Vargha-Delaney statistic ($\hat{A}$)~\cite{DBLP:journals/stvr/ArcuriB14}.

\subsection{\med size levels}
As can be seen from Figure~\ref{fig:Coverage-and-Time} (top row), both entity and entity connection coverage values tend to be quite high for all \med size levels, except for level S1 where the baseline performs poorly. For levels S2 and S3, the baseline achieves high coverage but not as high as \approach. Looking at Table~\ref{tab:Statistics}, we can see that for the three levels (S1, S2, S3) the differences between the coverage achieved by \approach and the baseline are statistically significant with a large effect size. In the other two levels (S4, S5) there is no statistically significant difference.

\begin{figure*}[!htb]
	\hspace*{\fill}%
	\subcaptionbox{\approach}{\includegraphics[width=0.48\textwidth]{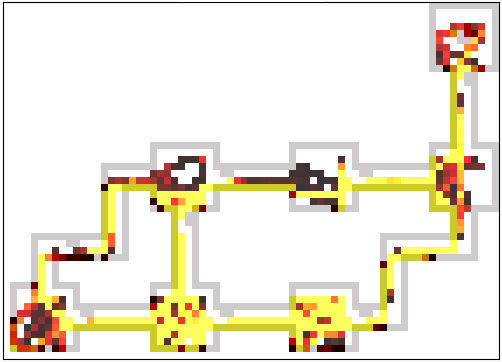}}\hfill%
	\subcaptionbox{baseline}{\includegraphics[width=0.48\textwidth]{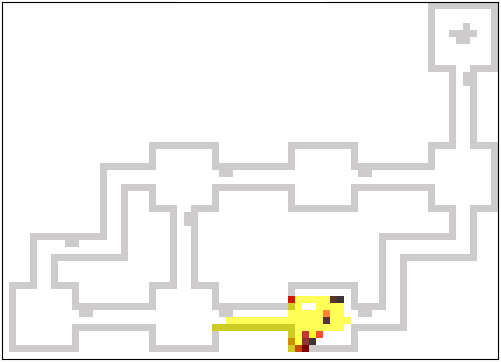}}%
	\hspace*{\fill}%
	\caption{Spatial coverage for level S1 of \med size. The darker the color the less explored the area.}
	\label{fig:MediumHeatmap}
\end{figure*}

Figure~\ref{fig:MediumHeatmap} shows the spatial coverage of one of the \med size levels (level S1 in Figure~\ref{fig:Coverage-and-Time}) where the stark contrast between the \approach and the baseline could be observed. We do not report all figures here due to lack of space, however, we make all data available in our online appendix~\cite{online-appendix}.

Looking at the time spent by both approaches on all of the levels (Figure~\ref{fig:Coverage-and-Time}; top row, last column), we can see that except level S1, where the baseline takes less time than \approach, in all the other cases \approach is in general faster than the baseline. The differences are also statistically significant (see Table~\ref{tab:Statistics}, 'Time' column) with the exception of S5 where the two approaches spend a comparable amount of time.

\subsection{\lrg size levels}
Differently from the \med size levels, the coverage achieved on the \lrg levels is generally lower (Figure~\ref{fig:Coverage-and-Time}, middle row). In particular, the baseline achieves low entity (below 60\%) and entity connection (below 50\%) coverage, while \approach achieves higher coverage (above 70\% entity and 50\% entity connection) but it is still low compared to the coverage achieved on the \med size levels.

\begin{figure*}[!htb]
	\hspace*{\fill}%
	\subcaptionbox{\approach}{\includegraphics[width=0.48\textwidth]{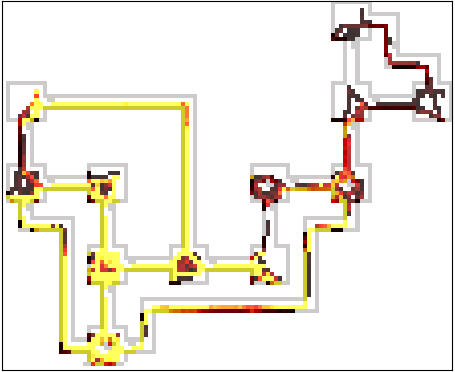}}\hfill%
	\subcaptionbox{baseline}{\includegraphics[width=0.48\textwidth]{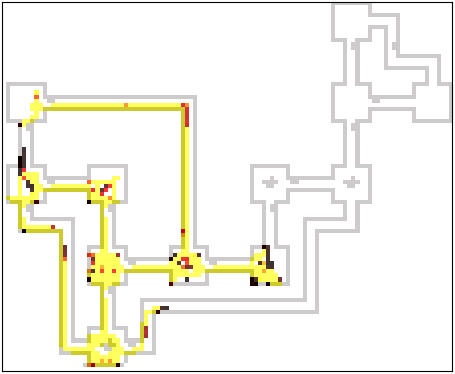}}%
	\hspace*{\fill}%
	\caption{Spatial coverage for level M3 of \lrg size. The darker the color the less explored the area.}
	\label{fig:LargeHeatmap}
\end{figure*}

Figure~\ref{fig:LargeHeatmap} shows the spatial coverage of one of the \lrg size levels (level M3 in Figure~\ref{fig:Coverage-and-Time}) where we observe the marked difference between the \approach and the baseline.

The total time spent for the three levels of \lrg size (Figure~\ref{fig:Coverage-and-Time}; middle row, last column) shows a similar trend as that of \med size levels where \approach is faster than the baseline, and in two cases (levels M1 and M3) the difference is statistically significant with large effect size (see Table~\ref{tab:Statistics}; middle row, 'Time' column). 

\subsection{\ext size levels}
As expected, the coverage achieved for levels of \ext level is lower than those of \med. As can be seen from Figure~\ref{fig:Coverage-and-Time} (bottom row), the maximum coverage achieved is around 55\%. Consistent with \med and \lrg levels, the coverage achieved by \approach is higher than that of the baseline. It is important to note that we have increased the budget allocated (number of episodes and number of actions per episode, see Table~\ref{tab:parameters}) since the levels are quite large. However the achieved coverage is low regardless, and hence future experimentation with larger budgets could be useful.

\begin{figure*}[!htb]
	\hspace*{\fill}%
	\subcaptionbox{\approach}{\includegraphics[width=0.48\textwidth]{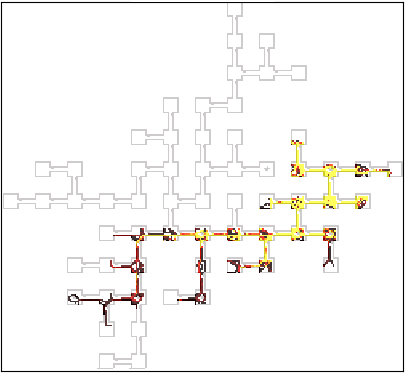}}\hfill%
	\subcaptionbox{baseline}{\includegraphics[width=0.48\textwidth]{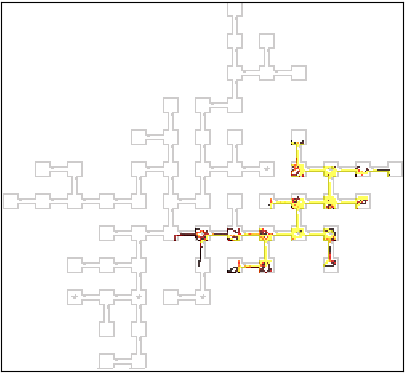}}%
	\hspace*{\fill}%
	\caption{Spatial coverage for level L3 of \ext size. The darker the color the less explored the area.}
	\label{fig:ExtremeHeatmap}
\end{figure*}

Figure~\ref{fig:ExtremeHeatmap} shows the spatial coverage for one of the \ext size levels (level L3 in Figure~\ref{fig:Coverage-and-Time}). As can be seen from the heatmap, the level is quite large and the agent (player) would need to cover a lot of ground to achieve full coverage.

Looking at the time spent for the four levels of \ext size (see Table~\ref{tab:Statistics}; bottom row, `Time' column), we observe a similar trend as the \med and \lrg size levels where in three of the levels (L1, L2, and L4) \approach is statistically significantly faster than the baseline where as for L3 there is no statistically significant difference in the time spent by the two approaches. 

\subsection{Discussion}
Overall the results show that \approach performs better than the single-agent variant, both in terms of coverage and time spent. Given the complexity of 3D games and the limitation due to partial observability, a single agent would need a lot of effort to effectively explore the game world. On the other hand, deploying multiple agents drastically improves performance since the active agent gets more accurate and immediate feedback on its actions. This tends to reduce the chances of the agent being `stuck' in certain areas of the game (see Figure~\ref{fig:MediumHeatmap}). Based on these results we can answer positively the first research question:
the \approach is \emph{more effective} than the single-agent variant.

Regarding the efficiency, i.e., time spent, the results are somehow unexpected in that \approach tends to consume less time compared to the baseline. Given the fact that two agents are running and exchanging information on actions performed, we had expected the time spent by \approach to be higher compared to the baseline. However, the results show the opposite. For the simpler levels (e.g., those of \med size) where \approach is able to achieve full coverage, the time spent is clearly less than the maximum allowed (since testing stops when full coverage is reached). On the other hand when full coverage is not achieved, it appears that in \approach the agent is less likely to get `stuck' and so less likely to waste more time trying to get out (see for example Figure~\ref{fig:MediumHeatmap}). Based on the results, for what concerns the second research question as well, we can answer positively that \approach is also \emph{more efficient} than the baseline.

\section{Related Works} \label{sec:relatedworks}

Recent years have seen remarkable progress in \rlacronym in solving immensely challenging multi-player real strategy video games \cite{vinyals2019grandmasterStarCraft,zhou2021hierarchical,OpenAIHideandSeekbaker2019,silver2017masteringGo}. The aim here is for an agent to achieve phenomenal skills to finish these games either by collaborating or competing with other agents. 
The promising result of \rlacronym in game play opens the possibility of using such solutions for automated testing of games. A few research work is found using single agent \rlacronym for game play testing and coverage~\cite{bergdahl2020augmenting, sestini2022automated, zheng2019wuji, holmgaard2018automated, ariyurek2019automated, tufano2022using, borovikov2019winning}.

Despite the growing success of \marlacronym, considering the inherent challenges of \marlacronym compared to its single agent counterpart, there is little work towards using \marlacronym architecture in automated game testing and coverage. 
One research work is found~\cite{gordillo2021improving}  
where the idea is to use multiple agents for maximizing game coverage. The solution uses multiple \rlacronym agents learning in a distributed fashion. Each agent is motivated by curiosity during learning. Achieving convergence in a distributed \marlacronym architecture is notoriously challenging, thus the authors do not focus on techniques for optimizing the training of agents in a distributed structure, rather they focus on collecting and combining the exploration data to measure the coverage. 
The use of \marlacronym in the field of automated game testing and coverage requires more research and our work goes in this direction where we show that employing two fully cooperating agents tends to achieve better coverage. Different agent configurations are the subject of future work.

\section{Conclusion} \label{sec:conclusion}
Automated testing of interactive software, such as 3D games, is challenging and has predominantly remained a manual task. In this paper, we have presented \approach, a curiosity driven \marlacronym approach for automated testing of 3D games. \approach adopts fully cooperative \marlacronym to explore the game world and achieve predefined coverage criteria. The experimental results show that \approach is effective and efficient compared to a baseline approach employing a single agent \rlacronym.

The work presented in this paper is an attempt to get insight into the use of multiple agents in the contest of testing 3D games. As such, it did not address all possible aspects related to the application of \marlacronym for testing 3D games. However, the results presented here serve as a good starting point for further exploring different dimensions. Hence, we envision a number of directions in which this work could be extended in future work. 

While in this work we adopted a cooperative \marlacronym scheme, other  agent schemes are worth exploring, such as a competitive scheme where agents compete against each other to achieve better performance. Moreover, fully cooperative agents where, unlike in this work, all agents are \emph{active} is also interesting to explore. In these cases, however, the environment could become non-stationary, which is not a trivial issue to manage. The number of agents deployed, beyond two, is another aspect that requires a comprehensive empirical exploration.

It is also worth investigating whether using deep \rlacronym could have an advantage over the tabular Q-learning algorithm used in this work, especially in cases where the size and complexity of the testing problem is higher. 

The current work mainly focused on achieving full coverage of the game level under test based on some predefined notions of coverage. It would be interesting to explore the possibility of identifying erratic behaviors that could result in a bug.

\section*{Acknowledgment}
We acknowledge the support  of the PNRR project FAIR -  Future AI Research (PE00000013),  under the NRRP MUR program funded by the NextGenerationEU.

\bibliographystyle{IEEEtran}
\bibliography{IEEEabrv,reference}

\begin{thebibliography}{10}
\providecommand{\url}[1]{#1}
\csname url@samestyle\endcsname
\providecommand{\newblock}{\relax}
\providecommand{\bibinfo}[2]{#2}
\providecommand{\BIBentrySTDinterwordspacing}{\spaceskip=0pt\relax}
\providecommand{\BIBentryALTinterwordstretchfactor}{4}
\providecommand{\BIBentryALTinterwordspacing}{\spaceskip=\fontdimen2\font plus
\BIBentryALTinterwordstretchfactor\fontdimen3\font minus
  \fontdimen4\font\relax}
\providecommand{\BIBforeignlanguage}[2]{{%
\expandafter\ifx\csname l@#1\endcsname\relax
\typeout{** WARNING: IEEEtran.bst: No hyphenation pattern has been}%
\typeout{** loaded for the language `#1'. Using the pattern for}%
\typeout{** the default language instead.}%
\else
\language=\csname l@#1\endcsname
\fi
#2}}
\providecommand{\BIBdecl}{\relax}
\BIBdecl

\bibitem{DBLP:conf/ast/PolitowskiPG21}
C.~Politowski, F.~Petrillo, and Y.~Gu{\'{e}}h{\'{e}}neuc, ``A survey of video
  game testing,'' in \emph{2nd {IEEE/ACM} International Conference on
  Automation of Software Test, AST@ICSE 2021, Madrid, Spain, May 20-21,
  2021}.\hskip 1em plus 0.5em minus 0.4em\relax {IEEE}, 2021, pp. 90--99.

\bibitem{silver2017masteringGo}
D.~Silver, J.~Schrittwieser, K.~Simonyan, I.~Antonoglou, A.~Huang, A.~Guez,
  T.~Hubert, L.~Baker, M.~Lai, A.~Bolton \emph{et~al.}, ``Mastering the game of
  go without human knowledge,'' \emph{nature}, vol. 550, no. 7676, pp.
  354--359, 2017.

\bibitem{mnih2013playing}
V.~Mnih, K.~Kavukcuoglu, D.~Silver, A.~Graves, I.~Antonoglou, D.~Wierstra, and
  M.~Riedmiller, ``Playing atari with deep reinforcement learning,''
  \emph{arXiv preprint arXiv:1312.5602}, 2013.

\bibitem{openAI-dota2}
C.~Berner, G.~Brockman, B.~Chan, V.~Cheung, P.~D{\k{e}}biak, C.~Dennison,
  D.~Farhi, Q.~Fischer, S.~Hashme, C.~Hesse \emph{et~al.}, ``Dota 2 with large
  scale deep reinforcement learning,'' \emph{arXiv preprint arXiv:1912.06680},
  2019.

\bibitem{DBLP:journals/tciaig/AlbaghajatiA23}
A.~Albaghajati and M.~A. Ahmed, ``Video game automated testing approaches: An
  assessment framework,'' \emph{{IEEE} Trans. Games}, vol.~15, no.~1, pp.
  81--94, 2023.

\bibitem{politowski2022towards}
C.~Politowski, Y.-G. Gu{\'e}h{\'e}neuc, and F.~Petrillo, ``Towards automated
  video game testing: still a long way to go,'' in \emph{Proceedings of the 6th
  International ICSE Workshop on Games and Software Engineering: Engineering
  Fun, Inspiration, and Motivation}, 2022, pp. 37--43.

\bibitem{DBLP:journals/tciaig/AspertiCPPS20}
\BIBentryALTinterwordspacing
A.~Asperti, D.~Cortesi, C.~D. Pieri, G.~Pedrini, and F.~Sovrano, ``Crawling in
  rogue's dungeons with deep reinforcement techniques,'' \emph{{IEEE} Trans.
  Games}, vol.~12, no.~2, pp. 177--186, 2020. [Online]. Available:
  \url{https://doi.org/10.1109/TG.2019.2899159}
\BIBentrySTDinterwordspacing

\bibitem{DBLP:conf/icst/PradaPKDVLDKDF20}
R.~Prada, I.~S. W.~B. Prasetya, F.~M. Kifetew, F.~Dignum, T.~E.~J. Vos,
  J.~Lander, J.~Donnart, A.~Kazmierowski, J.~Davidson, and P.~M. Fernandes,
  ``Agent-based testing of extended reality systems,'' in \emph{13th {IEEE}
  International Conference on Software Testing, Validation and Verification,
  {ICST} 2020, Portugal, October 24-28, 2020}.\hskip 1em plus 0.5em minus
  0.4em\relax {IEEE}, 2020.

\bibitem{DBLP:conf/kbse/FerdousKPS22}
R.~Ferdous, F.~M. Kifetew, D.~Prandi, and A.~Susi, ``Towards agent-based
  testing of 3d games using reinforcement learning,'' in \emph{37th {IEEE/ACM}
  International Conference on Automated Software Engineering, {ASE} 2022,
  Rochester, MI, USA, October 10-14, 2022}.\hskip 1em plus 0.5em minus
  0.4em\relax {ACM}, 2022.

\bibitem{prasetya2020aplib}
I.~Prasetya, M.~Dastani, R.~Prada, T.~E. Vos, F.~Dignum, and F.~Kifetew,
  ``Aplib: Tactical agents for testing computer games,'' in \emph{International
  Workshop on Engineering Multi-Agent Systems (EMAS)}.\hskip 1em plus 0.5em
  minus 0.4em\relax Springer, 2020, pp. 21--41.

\bibitem{OpenAIHideandSeekbaker2019}
\BIBentryALTinterwordspacing
B.~Baker, I.~Kanitscheider, T.~Markov, Y.~Wu, G.~Powell, B.~McGrew, and
  I.~Mordatch, ``Emergent tool use from multi-agent autocurricula,'' in
  \emph{International Conference on Learning Representations}, 2020. [Online].
  Available: \url{https://openreview.net/forum?id=SkxpxJBKwS}
\BIBentrySTDinterwordspacing

\bibitem{vinyals2019grandmasterStarCraft}
O.~Vinyals, I.~Babuschkin, W.~M. Czarnecki, M.~Mathieu, A.~Dudzik, J.~Chung,
  D.~H. Choi, R.~Powell, T.~Ewalds, P.~Georgiev \emph{et~al.}, ``Grandmaster
  level in {S}tar{C}raft {II} using multi-agent reinforcement learning,''
  \emph{Nature}, vol. 575, no. 7782, pp. 350--354, 2019.

\bibitem{park2019multi}
Y.~J. Park, Y.~S. Cho, and S.~B. Kim, ``Multi-agent reinforcement learning with
  approximate model learning for competitive games,'' \emph{PloS one}, vol.~14,
  no.~9, p. e0222215, 2019.

\bibitem{zhou2021hierarchical}
W.~J. Zhou, B.~Subagdja, A.-H. Tan, and D.~W.-S. Ong, ``Hierarchical control of
  multi-agent reinforcement learning team in real-time strategy ({RTS})
  games,'' \emph{Expert Systems with Applications}, vol. 186, 2021.

\bibitem{10.1145/3643658.3643920}
\BIBentryALTinterwordspacing
R.~Coppola, T.~Fulcini, S.~Manzi, and F.~Strada, ``How to measure game testing:
  a survey of coverage metrics,'' in \emph{Proceedings of the ACM/IEEE 8th
  International Workshop on Games and Software Engineering}, ser. GAS
  '24.\hskip 1em plus 0.5em minus 0.4em\relax New York, NY, USA: Association
  for Computing Machinery, 2024, p. 15–19. [Online]. Available:
  \url{https://doi.org/10.1145/3643658.3643920}
\BIBentrySTDinterwordspacing

\bibitem{canese2021multi}
L.~Canese, G.~C. Cardarilli, L.~Di~Nunzio, R.~Fazzolari, D.~Giardino, M.~Re,
  and S.~Span{\`o}, ``Multi-agent reinforcement learning: A review of
  challenges and applications,'' \emph{Applied Sciences}, vol.~11, no.~11,
  2021.

\bibitem{sutton2018reinforcement}
R.~S. Sutton and A.~G. Barto, \emph{Reinforcement learning: An
  introduction}.\hskip 1em plus 0.5em minus 0.4em\relax MIT press, 2018.

\bibitem{bellman1957markovian}
R.~Bellman, ``A markovian decision process,'' \emph{Journal of mathematics and
  mechanics}, pp. 679--684, 1957.

\bibitem{littman1994markov}
M.~L. Littman, ``Markov games as a framework for multi-agent reinforcement
  learning,'' in \emph{Machine learning proceedings 1994}.\hskip 1em plus 0.5em
  minus 0.4em\relax Elsevier, 1994, pp. 157--163.

\bibitem{DBLP:journals/corr/abs-2011-00583}
\BIBentryALTinterwordspacing
Y.~Yang and J.~Wang, ``An overview of multi-agent reinforcement learning from
  game theoretical perspective,'' \emph{CoRR}, vol. abs/2011.00583, 2020.
  [Online]. Available: \url{https://arxiv.org/abs/2011.00583}
\BIBentrySTDinterwordspacing

\bibitem{zhang2021multi}
K.~Zhang, Z.~Yang, and T.~Ba{\c{s}}ar, ``Multi-agent reinforcement learning: A
  selective overview of theories and algorithms,'' \emph{Handbook of
  reinforcement learning and control}, pp. 321--384, 2021.

\bibitem{DBLP:journals/corr/abs-1906-04737}
\BIBentryALTinterwordspacing
G.~Papoudakis, F.~Christianos, A.~Rahman, and S.~V. Albrecht, ``Dealing with
  non-stationarity in multi-agent deep reinforcement learning,'' \emph{CoRR},
  vol. abs/1906.04737, 2019. [Online]. Available:
  \url{http://arxiv.org/abs/1906.04737}
\BIBentrySTDinterwordspacing

\bibitem{gerstgrasser2024selectively}
M.~Gerstgrasser, T.~Danino, and S.~Keren, ``Selectively sharing experiences
  improves multi-agent reinforcement learning,'' \emph{Advances in Neural
  Information Processing Systems}, vol.~36, 2024.

\bibitem{DBLP:journals/tsmc/BusoniuBS08}
\BIBentryALTinterwordspacing
L.~Busoniu, R.~Babuska, and B.~D. Schutter, ``A comprehensive survey of
  multiagent reinforcement learning,'' \emph{{IEEE} Trans. Syst. Man Cybern.
  Part {C}}, vol.~38, no.~2, pp. 156--172, 2008. [Online]. Available:
  \url{https://doi.org/10.1109/TSMCC.2007.913919}
\BIBentrySTDinterwordspacing

\bibitem{pathak2017curiosity}
D.~Pathak, P.~Agrawal, A.~A. Efros, and T.~Darrell, ``Curiosity-driven
  exploration by self-supervised prediction,'' in \emph{International
  conference on machine learning}.\hskip 1em plus 0.5em minus 0.4em\relax PMLR,
  2017, pp. 2778--2787.

\bibitem{jaccard1912distribution}
P.~Jaccard, ``The distribution of the flora in the alpine zone. 1,'' \emph{New
  phytologist}, vol.~11, no.~2, pp. 37--50, 1912.

\bibitem{online-appendix}
\BIBentryALTinterwordspacing
R.~Ferdous, F.~M. Kifetew, D.~Prandi, and A.~Susi, ``{C}uriosity {D}riven
  {M}ulti-agent {R}einforcement {L}earning for {3D} {G}ame {T}esting, online
  appendix,'' FBK, Tech. Rep.~v1, 2025. [Online]. Available:
  \url{https://zenodo.org/doi/10.5281/zenodo.13886055}
\BIBentrySTDinterwordspacing

\bibitem{DBLP:journals/stvr/ArcuriB14}
\BIBentryALTinterwordspacing
A.~Arcuri and L.~C. Briand, ``A hitchhiker's guide to statistical tests for
  assessing randomized algorithms in software engineering,'' \emph{Softw. Test.
  Verification Reliab.}, vol.~24, no.~3, pp. 219--250, 2014. [Online].
  Available: \url{https://doi.org/10.1002/stvr.1486}
\BIBentrySTDinterwordspacing

\bibitem{bergdahl2020augmenting}
J.~Bergdahl, C.~Gordillo, K.~Tollmar, and L.~Gissl{\'e}n, ``Augmenting
  automated game testing with deep reinforcement learning,'' in \emph{2020 IEEE
  Conference on Games (CoG)}.\hskip 1em plus 0.5em minus 0.4em\relax IEEE,
  2020, pp. 600--603.

\bibitem{sestini2022automated}
A.~Sestini, L.~Gissl{\'e}n, J.~Bergdahl, K.~Tollmar, and A.~D. Bagdanov,
  ``Automated gameplay testing and validation with curiosity-conditioned
  proximal trajectories,'' \emph{IEEE Transactions on Games}, 2022.

\bibitem{zheng2019wuji}
Y.~Zheng, X.~Xie, T.~Su, L.~Ma, J.~Hao, Z.~Meng, Y.~Liu, R.~Shen, Y.~Chen, and
  C.~Fan, ``Wuji: Automatic online combat game testing using evolutionary deep
  reinforcement learning,'' in \emph{2019 34th IEEE/ACM International
  Conference on Automated Software Engineering (ASE)}.\hskip 1em plus 0.5em
  minus 0.4em\relax IEEE, 2019, pp. 772--784.

\bibitem{holmgaard2018automated}
C.~Holmg{\aa}rd, M.~C. Green, A.~Liapis, and J.~Togelius, ``Automated
  playtesting with procedural personas through mcts with evolved heuristics,''
  \emph{IEEE Transactions on Games}, vol.~11, no.~4, pp. 352--362, 2018.

\bibitem{ariyurek2019automated}
S.~Ariyurek, A.~Betin-Can, and E.~Surer, ``Automated video game testing using
  synthetic and human-like agents,'' \emph{IEEE Transactions on Games}, 2019.

\bibitem{tufano2022using}
R.~Tufano, S.~Scalabrino, L.~Pascarella, E.~Aghajani, R.~Oliveto, and
  G.~Bavota, ``Using reinforcement learning for load testing of video games,''
  in \emph{Proceedings of the 44th International Conference on Software
  Engineering}, 2022, pp. 2303--2314.

\bibitem{borovikov2019winning}
I.~Borovikov, Y.~Zhao, A.~Beirami, J.~Harder, J.~Kolen, J.~Pestrak, J.~Pinto,
  R.~Pourabolghasem, H.~Chaput, M.~Sardari \emph{et~al.}, ``Winning isn’t
  everything: Training agents to playtest modern games,'' in \emph{AAAI
  Workshop on Reinforcement Learning in Games}, 2019.

\bibitem{gordillo2021improving}
C.~Gordillo, J.~Bergdahl, K.~Tollmar, and L.~Gissl{\'e}n, ``Improving
  playtesting coverage via curiosity driven reinforcement learning agents,'' in
  \emph{2021 IEEE Conference on Games (CoG)}.\hskip 1em plus 0.5em minus
  0.4em\relax IEEE, 2021, pp. 1--8.

\end{thebibliography}

\end{document}